\documentclass[aps,prb,twocolumn,superscriptaddress,showpacs,nofootinbib,nolongbibliography,10pt]{revtex4-2}

\usepackage[utf8]{inputenc}
\usepackage{comment}
\usepackage{xcolor}
\usepackage{tikz}
\usepackage{times}
\usepackage[normalem]{ulem}
\usepackage{hyperref}
\usepackage{graphicx}

\newcommand{\gw}{$GW$}

\newcommand{\vect}[1]{\mathbf{#1}}
\newcommand{\vek}{\mathbf{k}}
\newcommand{\ver}{\mathbf{r}}

\newcommand{\lpt}{Laboratoire de Physique Th\'eorique, Universit\'e de Toulouse, CNRS, UPS, F-31062 Toulouse, France and European Theoretical Spectroscopy Facility (ETSF)}

\newcommand{\san}{Istituto di Struttura della Materia, Consiglio Nazionale delle Ricerche (CNR-ISM), Division of Ultrafast Processes in Materials (FLASHit), Via Salaria Km 29.5, CP 10, I-00016 Monterotondo Stazione, Italy}
\newcommand{\al}{Institute of Materials Science (ICMUV), Universitat de Val\`{e}ncia, Catedr\'{a}tico Beltr\'{a}n 2, E-46980, Valencia, Spain}

\begin{document}
\title{Assignment of excitonic insulators in \textit{ab initio} theories: the case of NiBr$_2$}

\author{Stefano Di Sabatino}
\email{stefano.disabatino@irsamc.ups-tlse.fr}
\affiliation{\lpt}
\author{Alejandro Molina-S\'{a}nchez}
\email{alejandro.molina@uv.es}
\affiliation{\al}
\author{Pina Romaniello}
\email{pina.romaniello@irsamc.ups-tlse.fr}
\affiliation{\lpt}
\author{Davide Sangalli}
\email{davide.sangalli@ism.cnr.it}
\affiliation{\san}

\begin{abstract}
In this work we perform a detailed first--principles analysis of the electronic and optical properties of NiBr$_2$ within the state--of--the--art \gw+BSE scheme to determine whether this system displays negative excitonic energies, which would identify it as an (half) excitonic insulator. Particular attention is payed to the convergence of the \gw band structure and to the consistency between approximations employed in the ground-state calculations and approximations employed in the linear response calculations. We show that these two issues play a crucial role in identifying the excitonic nature of NiBr$_2$.
\end{abstract}

\maketitle

\section{Introduction}

The seminal paper~\cite{Jerome1967} of Jerome and coworkers predicted the realization of the broken symmetry excitonic insulator (EI) phase in materials where the exciton binding energy overcomes the fundamental band gap. The phase transition from a symmetric state to the lower-symmetry EI would be purely electronically driven. A very rich activity followed, with various research groups trying to find signature of EI phase in different class of materials, ranging from semi-metals to small gap semiconductors and nanostructures\cite{Bucher1991,Bronold2006,Cercellier2007,Su2008,Wakisaka2009,Ataeie2010,Varsano2017,Hellgren2018,Dong2020,Jiang2020,Mazza2020,Baldini2020,Sun2022,Chen2022}.

The determination of the EI phase is mostly quantitative, rather then qualitative, i.e. it relies on the relative value of the excitonic binding energy versus the fundamental band gap. Therefore accurate numerical simulations are essential. Moreover, the broken symmetry phase is often associated to an instability of the crystal structure, and a distortion towards a lower symmetry phase. It then becomes a chicken and egg question whether it is the lattice distortion that drives the electronic phase transition or vice-versa, and whether or not one should speak about excitonic instability at all in many cases. The structural and the electronic phase transitions cannot be disentangled experimentally, making a direct proof of the EI existence a non trivial problem, with different experimental groups often reaching opposite conclusions. This is why \textit{ab initio} numerical simulations are the key instrument to be used. Indeed, while indications of the EI phase from experimental measurements have been collected, in most works the assignment of the EI phase is based on \textit{ab initio} numerical simulations, either as support of experimental data or as purely theoretical/computational works.

The state-of-the-art approach to model excitons in \textit{ab initio} simulations is the so-called Bethe-Salpeter equation (BSE), computed on top of an accurate band structure including quasi-particle \gw corrections, within the so-called \gw+BSE scheme. Then, numerically, a material is foreseen to be an excitonic insulator if the BSE binding energy overcomes the \gw band gap. This approach has been used in several works.\cite{Ataeie2010,Varsano2017,Jiang2019,Jiang2020,Dong2020}
Recently, following this approach, a possible EI phase was identified by Jiang \textit{et al.}\cite{Jiang2019} for the magnetic layered NiBr$_2$.
NiBr$_2$ is quite unique for two reasons. First of all, it has a very large band gap (about 4 eV), compared to other materials explored so far. This would suggest that the EI phase could be ruled out from the beginning, since typical exciton binding energies range from few meV to few hundreds meV in bulk materials. Second, NiBr$_2$ is magnetic, thus leading to an interplay between charge and spin degrees of freedom. As a result, the EI phase was predicted to be realized in one spin channel only. The authors referred to this case as an half-EI (HEI).

The \gw+BSE scheme is a well established numerical scheme, and it works surprisingly well in describing the optical properties of a wide range of materials, ranging from bulk insulators and semi-conductors, to layered and 2D materials, 1D materials, such as carbon nanotubes, up to isolated systems such as clusters and molecules.
In NiBr$_2$, similarly to other EI candidates, the \gw+BSE scheme results in solutions of the excitonic hamiltonian with slightly negative energies (few meV below zero). However the validity of the \gw+BSE scheme to correctly capture poles with energy close, or even below, zero should be questioned from the beginning for various reasons. 
First, it is well known that the results of the \gw+BSE scheme depend on the starting point, i.e. on the functional employed in the initial density functional theory (DFT) calculation. For example, if the initial band structure is computed starting from a DFT calculation using the PBE or the PBE0 approximation, one should refer to it as \gw@PBE+BSE or \gw@PBE0+BSE, respectively. Although a change of the functional might only result in small energy shifts, this could be enough to move an energy eigenvalue of the BSE across zero.
Second, in this protocol approximations employed in the ground-state calculations are not consistent with the approximations used in the linear response part. As a consequence, for example, the \gw@DFT+BSE scheme, if used to compute excitations in the spin flip channel, fails in reproducing the Goldstone sum rule for magnons, i.e. it fails to give a zero energy pole in magnetic materials~\cite{2017Muller_PhD}.
This failure is independent of the DFT flavour, the corresponding error can be large, and it clearly shows that the scheme is not reliable for poles with energy close to zero.

Surprisingly, these issues have been discussed very little in the literature when addressing the existence of possible EI phases.
In the present manuscript we use the quite unique case of NiBr$_2$ to address this deficiency. We show that NiBr$_2$ is an extreme case where one should be very careful in proposing the existence of an EI phase, since the state-of-the-art \gw@DFT+BSE scheme is so much affected by the starting point, that even qualitatively results can change completely. These findings are not specific to the case of NiBr$_2$ but can be extended to the analysis of a wide range of materials.

The paper is organized as follows. In Sec.~\ref{Sec:Theory} we define in a clear way what an EI phase (or an EI instability) is, and discuss the numerical protocols which properly address the existence of the EI phase in general. In particular we discuss why one has to be consistent in the approximations used to describe the ground state and the response of the system.  In Sec.~\ref{Sec:Results} we  present the results of our numerical simulations and discuss the band structure (subsection \ref{Subsec:ResultsGW}) as well as the absorption spectrum (subsection \ref{Subsec:ResultsBSE}) of ferromagnetic NiBr$_2$ using various approximations. We finally draw our conclusions and perspectives in Sec.~\ref{Sec:Conclusions}.

\section{Theoretical protocol\label{Sec:Theory}}

The EI phase has two fundamental key properties: (i) it is a broken symmetry phase of electronic nature, and (ii) it is described by introducing the long range electron-hole interaction in the simulations. These are the two essential requirements that any numerical simulation predicting an EI phase should take into account.

Let us define the electronic hamiltonian $H$, which also includes the potential of the atoms in a chosen crystal lattice structure, and its eigenvalues (at fixed electron number) $E_\lambda$, with $E_0$ the ground-state energy. We can define a corresponding point symmetry group $S$ and a group of discrete translations $T$, such that $[H,S]=0$ and $[H,T]=0$. We call $\Omega^{ST}$ the phase space of electronic states which respect such symmetry groups, $\Omega$ the global phase space, and $\Omega^{S}$ ($\Omega^{T}$) the phase space which respects the $S$ ($T$) group alone. We have $\Omega \supseteq \Omega^{S} \supseteq \Omega^{ST}$ ($\Omega \supseteq \Omega^{T} \supseteq \Omega^{ST}$). The excitonic insulator phase arises from a spontaneous symmetry breaking, usually with respect to the $S$ group, and will be the focus of the present manuscript. 
Accordingly, we define $E^{ST}_0$ as the global energy minimum over $\Omega^{ST}$, and $E^{T}_0$ as the global energy minimum over $\Omega^{T}$. A broken-symmetry phase ground state exists if and only if $E^{T}_0<E_0^{ST}$.

Standard \textit{ab initio} simulations compute the ground state with the symmetry specified by the crystal lattice structure of the system, within a given approximation for the exchange and correlation (xc) energy; we indicate the corresponding total energy as $E_0^{ST,\text{xc}_g}$, with $\text{xc}_g$ referring to the xc approximation used in the ground state. There are then two options available to check if a broken-symmetry state with energy $E_0^{T,\text{xc}_g}<E_0^{ST,\text{xc}_g}$ exists:
(i.1) to perform a second ground-state simulation without symmetry or (i.2) to compute the spectrum of the neutral excitations of the material, $\omega_\lambda=(E_\lambda-E_0)$, at zero transferred momentum.\footnote{The finite momentum excitations would describe the breaking of the transnational symmetries as well. We do not consider this case here.} In the latter case there is no need to impose symmetry breaking. 
Indeed, neutral excitations are computed from the solution of the linear response function $\chi(\omega)$. The excited-state energies, $E_\lambda$, and the corresponding wavefunctions, $\Psi_\lambda$, which appear in the Lehmann representation of $\chi(\omega)$, do not belong to $\Omega_{ST}$ in general. More precisely, given the space symmetry group $S$, the states $\Psi_\lambda$ can be classified as degenerate multiplets of the irreducible representations $R_S^\lambda$ of the space group. Symmetry operations of $\Omega_{ST}$ can in general send a state of a given multiplet into any other state of the same multiplet.
Because of that, the states of a monodimensional representation $R_S^\lambda$ are singlets, i.e. nondegenerate, and belong to $\Omega_{ST}$. Instead, states of a multidimensional representations $R_S^I$ do not belong to $\Omega_{ST}$,
and they must be considered together to obtain a closed sub-space with respect to the symmetry operation of the group $S$. If a multiplet of states has energy lower than the ground state, then a spontaneous symmetry breaking would be favored, with an emerging ground state composed by a random linear combination of the states in the multiplet. 

If option (i.1) is chosen, the same level of approximation must be used in the ground-state calculation with and without symmetry. Comparing $E_0^{T,\text{xc}_{g1}}$ and $E_0^{ST,\text{xc}_{g2}}$, with $\text{xc}_{g1}$ and $\text{xc}_{g2}$ two different xc approximations used in the ground state, does not have any validity. Moreover, the long range electron-hole interaction needs to be considered. In some works the EI phase has been predicted on the basis of calculations using the local density approximation (LDA) or the generalized gradient approximation (GGA), i.e. $\text{xc}_g= \text{LDA/GGA}$. While also within these approximations a symmetry-broken phase could be obtained, such phases should not be labelled as EI, since \text{LDA/GGA} completely miss the long-range behaviour of the electron--hole interaction. Indeed, it is well known that time dependent DFT (TDDFT), within the LDA or GGA approximations (TDLDA or TDGGA), cannot describe excitons in the excited states, and thus these approximations cannot capture the EI phase in the ground state either.

In case option (i.2) is chosen, the total energy of possible broken-symmetry phases can be obtained by defining $E_\lambda^{T,\text{xc}}=E_0^{ST,\text{xc}_g}+\omega^{ST,\text{xc}_e}_\lambda$, with $\omega^{ST,\text{xc}}_\lambda=E_\lambda^{T,\text{xc}}-E_0^{ST,\text{xc}}$ a zero-momentum neutral excitation of the system, using consistent approximations between ground state ($\text{xc}_g$) and excited state ($\text{xc}_e$) calculations. Then the criteria to establish the existence, for some $\lambda_0$, of a broken-symmetry phase becomes $\omega^{ST,\text{xc}_e}_{\lambda_0}<0$. In such situation we can identify $E_{\lambda_0}^{T,\text{xc}}=E_0^{T,\text{xc}}$ as the global ground state energy.
The state-of-the-art scheme to describe excitons in the \textit{ab initio} framework is BSE. $\omega^{\text{xc}_e}_\lambda<0$, with ${\text{xc}}_e$ referring to the xc approximation used for the BSE kernel, is then often used as the golden standard to predict the existence of the excitonic insulator phase. Within this approach the \text{essential} consistency condition at the core of option (i.2) would rely on the use of a one-body Green’s function which is a self-consistent
solution of the Dyson equation for a given self-energy approximation and on the use of the same level of approximation for the self-energy and the BSE kernel \cite{2017Muller_PhD}.

However, in practical calculations, this consistency condition is violated since the state-of-the-art is the \gw@DFT+BSE scheme. Indeed, the BSE kernel is obtained from the Hartree plus screened-exchange self-energy (HSEX) approximation, i.e. $\text{xc}_e=\text{HSEX}$, while $\text{xc}_g=GW\text{@DFT}$. The self-consistent DFT ground state with \gw\, corrections should be replaced by a self-consistent simulation with $\text{xc}_g=\text{HSEX}$. Even ignoring this point and assuming that \gw\, is a good starting point for a BSE calculation, at least an eigenvalue self-consistent \gw\, (ev\gw) procedure should be carried out.
However both SC-HSEX and ev\gw\, can be very demanding. An alternative path is to use DFT with hybrid functionals, which are computationally affordable. Hybrid functional simulations include (a fraction of) the long--range electron--hole interaction and can be used to capture EI phases. In this case  $\text{xc}_g=\text{xc}_e=\text{Hybrid}$ can be used, by employing TDDFT for the excited states. To underline this consistency, we will use here the notation PBE0+TDPBE0 for example for the PBE0 case. Indeed, it has been shown that hybrid functionals can capture excitonic excited states~\cite{Joachim2008} and EI phases~\cite{Hellgren2018}. In general the exchange fraction entering hybrids is not properly screened, and this could be particularly problematic in 2D materials, since the 2D nature of the screening is completely missing. Nevertheless, as we shall see, it remains a valid approach to demonstrate the concept of consistency we discussed above.

In this work we will adopt the protocol described in option (i.2) to identify possible EI phases in NiBr$_2$ and we will calculate the optical spectrum within \gw@PBE+BSE, \gw@PBE0+BSE and PBE0+TDPBE0. 
This will allow us to address both the starting-point dependence of the \gw@DFT+BSE scheme and to consider a proper scheme with $\text{xc}_g=\text{xc}_e$.

\section{Results \label{Sec:Results}}

We studied NiBr$_2$ in its ferromagnetic ground state. 
Density functional theory calculations were performed using the Quantum
Espresso code~\cite{Giannozzi_2009}  within both the Perdew-Burke-Ernzerhof (PBE) and the PBE0 exchange correlation functional~\cite{Perdew1996}.
Optimized norm-conserving Vanderbilt pseudopotentials  have been generated using ONCVPSP \cite{Hamann2013}. 
The self-consistent (scf) simulations converged with an energy cutoff of 80 Ry and a $\vect{k}$-point grid of $9\times9\times1$ (the direct band gap at $\Gamma$ has been converged to 0.01 eV).

We then used the DFT wave-functions $\psi^{KS}_{n\vek}(\ver)$ and energies  $\epsilon^{KS}_{n\vek}$ to perform \gw\, and BSE simulations. For the PBE case we re-computed $\psi^{KS}_{n\vek}(\ver)$ and  $\epsilon^{KS}_{n\vek}$ with a non self-consistent (nscf) calculation on a $19\times19\times1$ $\vect{k}$-point grid in the Brillouin zone. For the PBE0 case instead we used a $9\times9\times1$ $\vect{k}$-point grid. This is due both to the higher computational cost of PBE0, and to the fact that, with hybrid functionals, it is not possible, in the nscf calculations, to use a $\vect{k}$-point grid different from the one used for the scf case. 
As we shall see, this restriction does not affect our conclusions. We computed the one-shot \gw quasiparticle (QP) corrections on top of DFT using the Yambo code \cite{Marini2009, Sangalli_2019}.
Dynamical screening effects have been taken into account using the Godby-Needs plasmon-pole approximation (PPA)~\cite{Godby1989}. 
Both in DFT and, successively, in $GW$ and BSE, we used a slab model with a 40 a.u. vacuum thickness.
Moreover, the Coulomb cut-off technique \cite{PhysRevB.73.205119} has been used to ensure no interactions between periodic images.
In the \gw step, we used a cutoff of 40 Ry for the wave-functions; an energy cut-off four times bigger, i.e. of 320 Ry, for $V_{\text{xc}}$ and for the exchange part of the Self-Energy $\Sigma_x$; finally an energy cutoff of 10 Ry for the screening and the correlation parts of the self-energy $\Sigma_c$ (see the Supplemental Material more for details \cite{supmat}). We included a total of 200 bands in the screening and 500 in the evaluation of the self-energy.
In the optical response we took into account excitonic effects  via
the Bethe-Salpeter Equation \cite{martin_reining_ceperley_2016, PhysRevB.62.4927}.
The BSE absorption spectrum has been calculated using a static screening converged using the same parameters of the \gw calculation. 
We used a cutoff of 4 Ry for  the electron-hole exchange and 10 Ry for the screened interaction. We included bands from 7 to 23 to obtain a converged spectrum up to 10 eV.

In the following we demonstrate that the occurrence of negative energy solutions of the BSE strongly depends on the choice of the approximations used in the ground state and for the BSE kernel. In particular the lack of self-consistency in the \gw\, band structure and of consistency between ground-state approximations and BSE kernel approximations can lead to a questionable EI phase.

\begin{figure*}[t]
    \centering
    \includegraphics[width=0.47\textwidth]{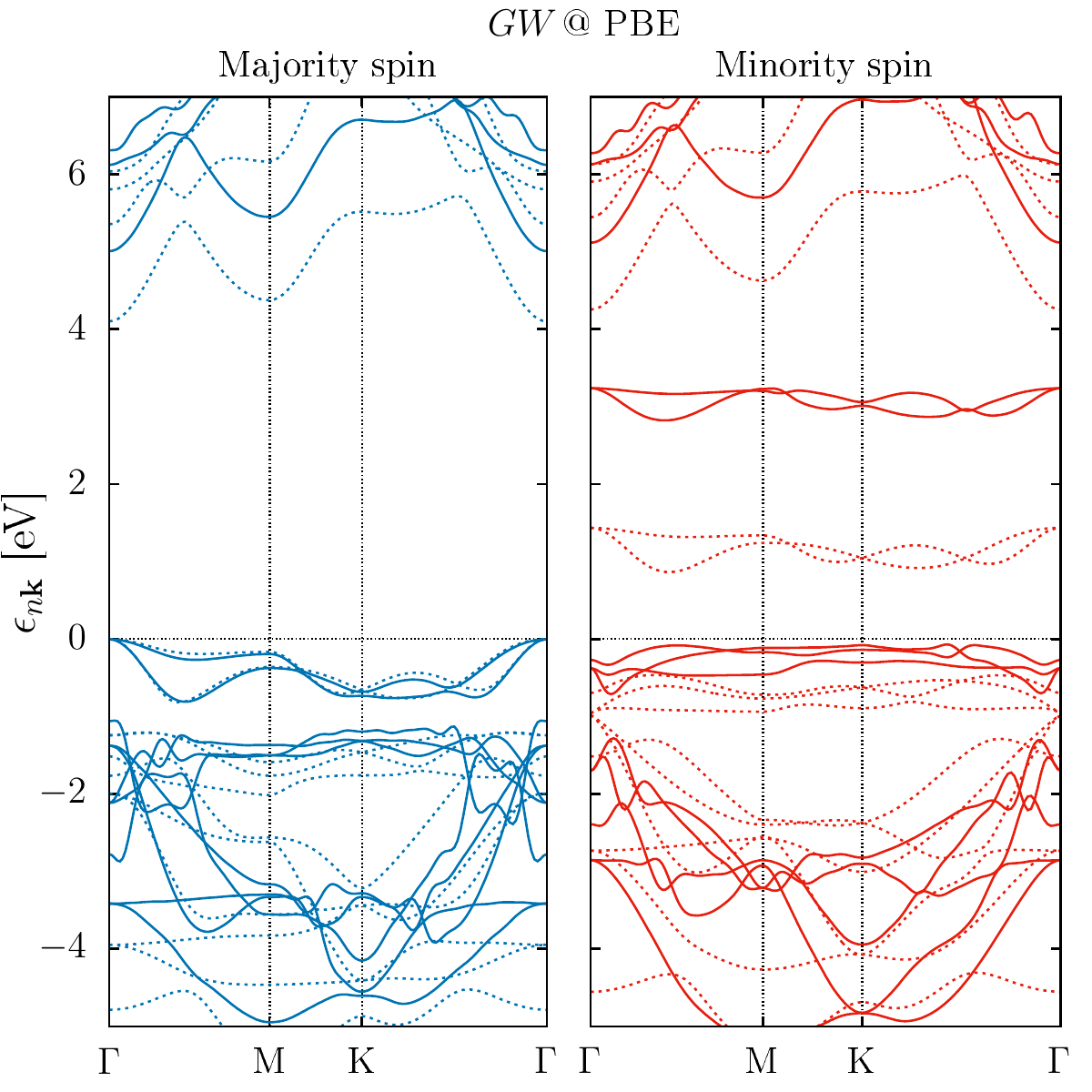}\hspace{8pt}
    \includegraphics[width=0.47\textwidth]{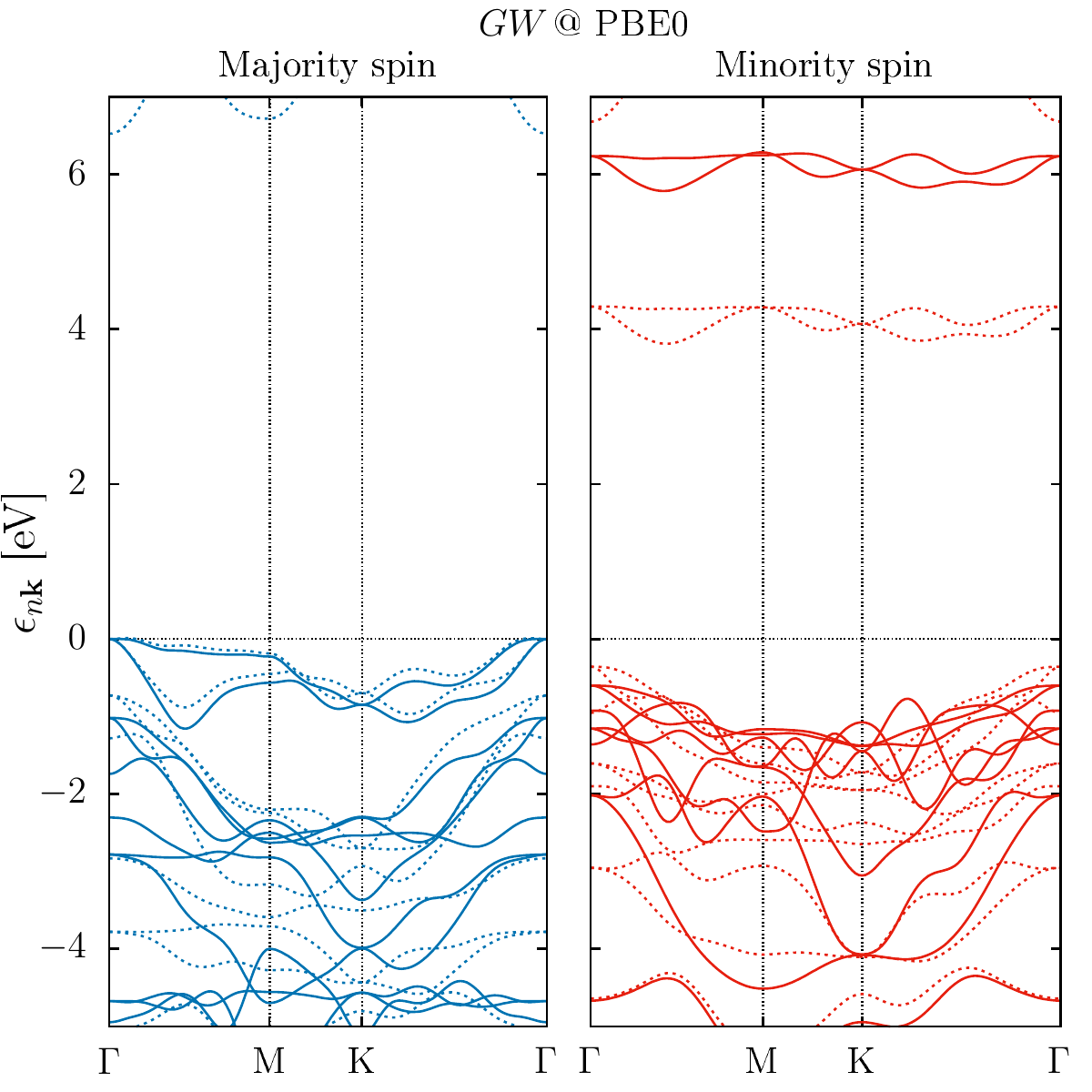}
     \caption{Spin-resolved band structure (majority- and minority-spin channels are reported in blue and red, respectively) for \gw@PBE (left) and \gw@PBE0 (right) are reported with a continuous line.
     The DFT band structures used as starting point for the  \gw correction are reported with a dashed line. The VBM is set to zero. 
     }
    \label{fig:bandGW}
\end{figure*}

\subsection{Band structure \label{Subsec:ResultsGW}}

The starting point for a BSE calculation is a single-particle band structure obtained correcting the DFT energies. The approximation chosen for the DFT determines the quality of the screened interaction $W$, which is computed within the RPA approximation using Kohn-Sham energies. This results in the so-called $W_0$ screened interaction. $W_0$ enters the one-shot \gw or $G_0W_0$ approach, which is usually employed for the band structure. The one-shot \gw is also employed in the present work.
Even without invoking self-consistency, \gw calculations can be problematic in 2D materials.
We found, indeed, that, for the system under study,  particular attention has to be paid when converging the \gw\, quasiparticle corrections. See also the discussion in the Supplemental Material \cite{supmat} (see, also, references \cite{guandalini2022efficient, c2db} therein).

\begin{table}[b]
    \centering
    \begin{tabular}{l|cc|c|cc|c}
    \hline\hline
         & \multicolumn{2}{c|}{PBE}   & PBE0 &  \multicolumn{3}{c}{\gw}
         \\
         \hline
         &  & \cite{c2db}  &  & @PBE  & @PBE \cite{c2db} & @PBE0
         \\
         \hline
    ind. gap M-m            & 0.83 & 0.88  & 3.79 & 2.82 & 2.62  & 5.75
    \\
    dir. gap m-m             & 1.33 & 1.34 & 4.45 & 2.91 
    & 2.62 & 6.56
    \\
    dir. gap M-M             & 4.09 & 3.95 & 6.52 & 5.01 & 5.37  & 7.41
    \\   
    dir. gap M-m             & 0.99 & 1.04 & 3.86 & 3.05 & 3.06  & 5.91
    \\    
    \hline\hline
    \end{tabular}
    \caption{Indirect and direct band gaps (in eV).}
    \label{tab:gap}
\end{table}

In Fig.~\ref{fig:bandGW} we show PBE and PBE0 spin-resolved band structures, separated in majority (M) and minority (m) spin channels. The corresponding \gw@PBE and \gw@PBE0 band structures are also shown. In all cases the band gap is indirect, with 0.83 eV at the PBE level compared to 3.79 eV at the PBE0 level, and it is due to transitions from the majority valence band maximum (VBM) at $\Gamma$ towards the minority conduction band minimum (CBM), which is at roughly half way along the $\Gamma$M high symmetry line. Also the direct band gaps within PBE0 (4.45 eV and 6.52 eV for m-m and M-M spin channels, respectively) are significantly larger than the ones obtained using PBE (1.33 eV and 4.09 eV for m-m and M-M spin channels, respectively). This is expected due to the fraction of HF exchange taken into account in the PBE0 functional.

Moving to \gw, one would naively expect that, despite the starting-point dependence, the \gw@PBE and \gw@PBE0 gaps should be close to each other, with \gw@PBE approaching the expected self-consistent \gw\, result from below, i.e. with the \gw\, corrections opening the PBE gap, and with \gw@PBE0 from above, i.e. with the \gw\, corrections  closing the PBE0 gap. However this is clearly not the case as shown in Fig.~\ref{fig:bandGW}. \gw@PBE opens the PBE gap from 0.83 to 2.62 eV. Direct band gaps are increased to 2.91 eV and 5.01 eV for the minority and majority spin channels, respectively. However \gw@PBE0 also opens the PBE0 gap from 3.79 to 5.75 eV. Direct band gaps are 6.56 eV and 7.41 eV for the minority and majority spin channels respectively. See also Tab.~\ref{tab:gap} for a comparison of all values.
This clearly points to the fact that $W_0^{\text{PBE}}$ and $W_0^{\text{PBE0}}$ are different and that, accordingly, the two $G_0W_0$ self-energies are different. One can then wonder which result is the most reliable. 

On one hand the \gw@PBE is the state-of-the-art scheme to compute quasi-particle band structure. Indeed, the PBE screening entering $W_0^{\text{PBE}}$ is usually a reasonable approximation to the physical screening due to error cancellations between (i) the underestimated fundamental gap, and (ii) the neglected excitonic effects. On the other hand hybrid functionals like PBE0 are expected to better approximate quasi-particle wave-functions and energies. Also simulations based on the hybrid HSE0 functional~\cite{Mushtaq2017,Jiang2019} give a band structure rather close to the PBE0 band-structure. However \gw@hybrids is a scheme which has not been much tested yet, since the compatibility of \gw\, codes with hybrid functionals is a quite recent feature, and, despite  being very promising~\cite{Fuchs2007,Lappert2019,Gant2022}, they are not very well tested.
By correcting the electronic gap, hybrid functionals tend to underestimate the electronic screening 
which may result in a overestimation of the \gw@hybrids gap~\cite{Fuchs2007}. Indeed, in NiBr$_2$ the underestimated screening leads to a very strong exchange term, even bigger than the exchange fraction entering in PBE0. As a consequence, it is more reasonable to directly compare \gw@PBE with hybrids like PBE0.
The estimated experimental band gap, probed using STS on NiBr$_2$ thin films, down to the monolayer, ranges between 3.4 and 4.5 eV, in good agreement with the PBE0 value~\cite{Bikaljevic2021}.


Beside this evident, but yet quantitative difference, there is a more important qualitative difference between PBE and PBE0 band structures, which is also inherited by the corresponding $GW$ band structure. Within PBE there exist three almost flat bands in the minority channel which are very close in energy to the top of the valence band in the majority spin channel. These bands are Nickel $3d$ states~\cite{Jiang2019}, in particular with $t_{2g}$ character, which are likely not well described within PBE. Indeed, they are not present within our PBE0 simulations, nor in the HSE0 and LDA+$U$ results in the literature~\cite{Mushtaq2017,Jiang2019}. Moving to \gw@PBE these same states remain there, pointing to the fact that the PBE wave-functions are not well described, and that PBE is not a good starting point. These states also show a quite peculiar behaviour in the \gw\, simulations, and, as we shall see, they are responsible for the very sharp difference in the spectrum of \gw@PBE+BSE compared to PBE0+TDPBE0 or \gw@PBE0+BSE. In the Supplemental Material \cite{supmat} we discuss the convergence of the \gw@PBE quasi-particle corrections. We highlight the ``unusual" behaviour of these flat Ni $t_{2g\downarrow}$ bands, and we discuss in detail the very slow convergence compared to the energy cut-off used in the description of the screened interaction.
The discussion agrees with previous numerical studies~\cite{Huser2013}, although this fact has been somehow overlooked in past simulations on 2D materials~\cite{Jiang2019}.
%
Indeed \gw@PBE corrections on these occupied Ni $t_{2g\downarrow}$ states lead to a situation in which the majority and the minority VBM are very close in energy (during the \gw convergence their relative position is even inverted). The direct gap switches character, from M-m to m-m, and the system almost becomes a direct gap material with an energy difference between \gw@PBE direct and indirect gap of 9 meV (compared to 500 meV in PBE, 660 meV in PBE0 and 810 meV in \gw@PBE0). The small m-m gap is also part of the reason why negative poles will appear in the response function calculated at the level of \gw@PBE+BSE, as we will discuss in Sec.~\ref{Subsec:ResultsBSE}.

The conclusion of this analysis is that NiBr$_2$ is an indirect band gap semiconductor. The indirect gap is majority-minority, as expected in standard magnetic materials. Looking at the direct gap, this is also majority-minority, the minority-minority gap is slightly larger (less than 1 eV), while the majority-majority gap is significantly larger. This behaviour is due to the empty spin-minority Ni $3d$ orbitals, with $e_g$ character, which lay in between the more predominantly $p$-like valence and conduction bands.

\subsection{Optical properties  \label{Subsec:ResultsBSE}}

\begin{figure}
    \centering
    \includegraphics[width=0.45\textwidth]{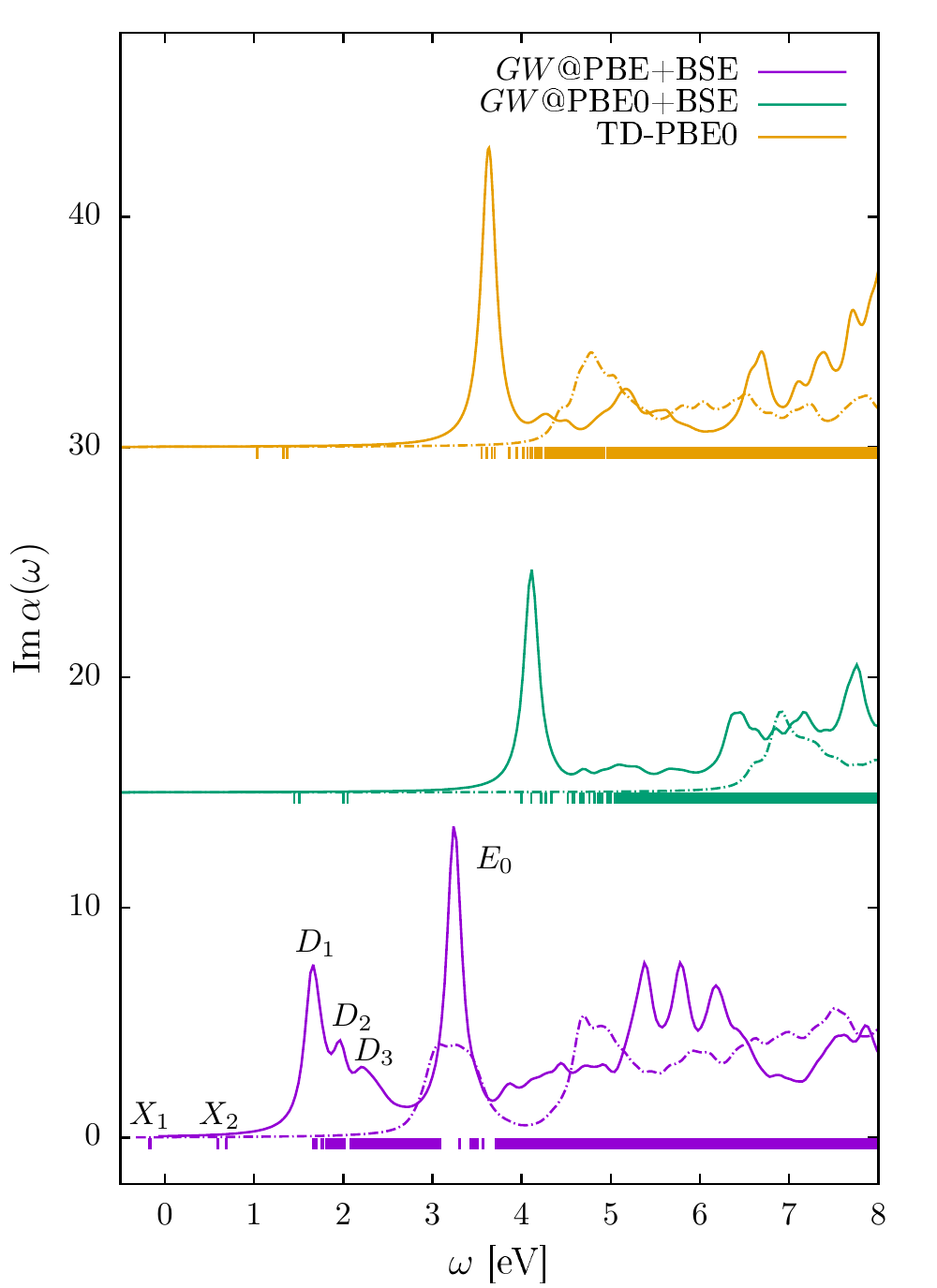}
    \caption{Imaginary part of the macroscopic polarizability calculated within 
    \gw@PBE+BSE (upper panel), \gw@PBE0+BSE (middle panel), and PBE0+TDPBE0 (lower panel). The results obtained within the corresponding independent particle approximation, i.e. ignoring
electron-hole interaction, are also reported (dashed lines). Curves are vertically shifted for clarity.
     Bright and dark excitons energies are reported as vertical thicks. }
    \label{fig:BSE}
\end{figure}

\begin{figure}[t]
    \centering
    \includegraphics[width=0.49\textwidth]{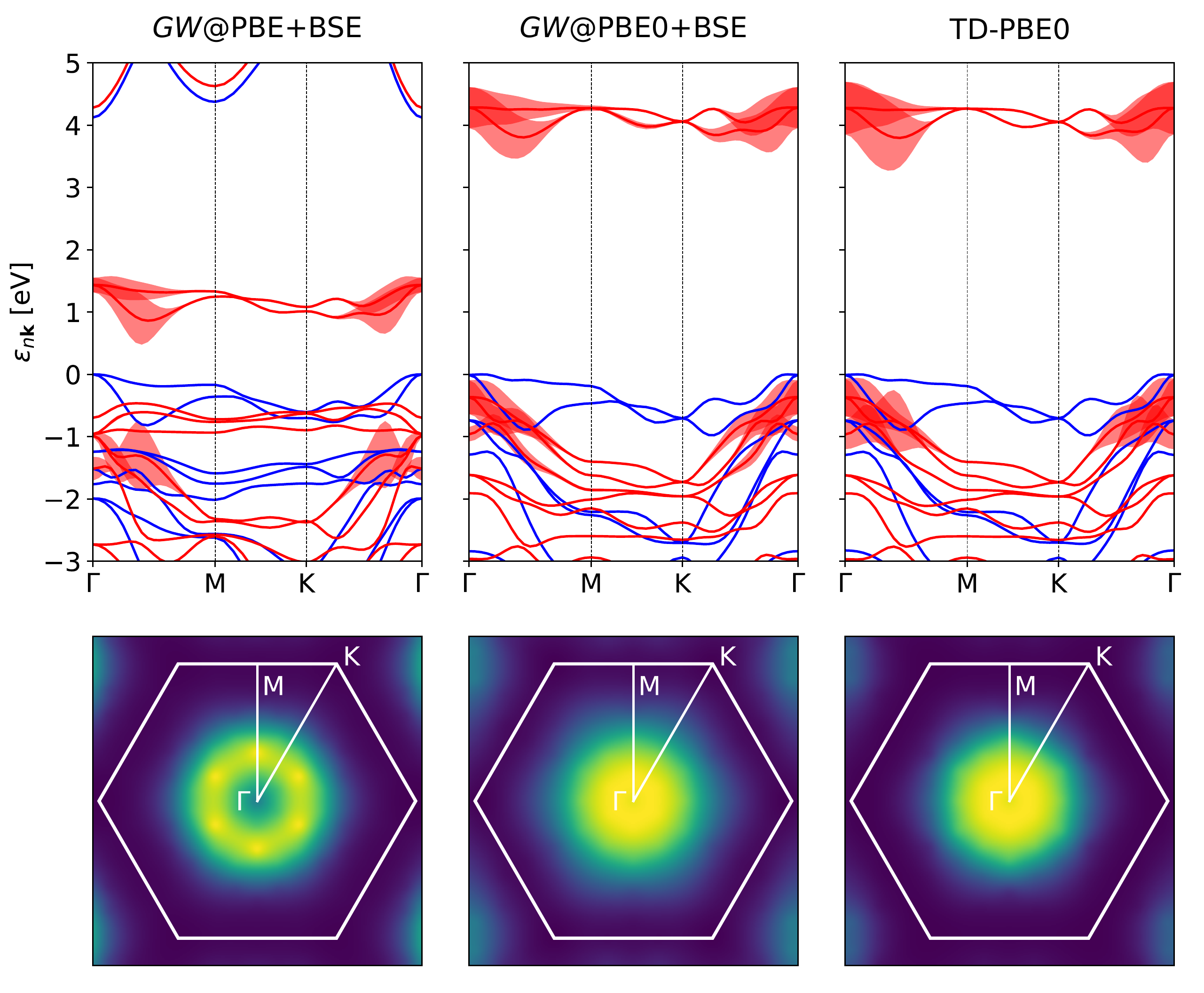}
    \caption{Band contribution to the main excitonic peak $E_0$: (left panel) \gw@PBE+BSE, (middle panel) \gw@PBE0+BSE, (right panel) PBE0+TDPBE0. Note that the plotted band structures are the DFT ones. We also report the excitonic wavefunction in reciprocal space. 
    }
    \label{fig:BSEcompExc}
\end{figure}

We now turn our attention to the optical properties of NiBr$_2$.
In Fig.~\ref{fig:BSE} the \gw@PBE+BSE absorption spectrum is reported together with the \gw@PBE0+BSE and the PBE0+TDPBE0 spectra.
All spectra are dominated by an intense excitonic peak, which we label $E_0$. All methods describe $E_0$ as an exciton mostly located nearby $\Gamma$ in the BZ  (see Fig.~\ref{fig:BSEcompExc}) and corresponding to a transition from $p$ bands $\epsilon_{p\vek\downarrow}$ to $d(e_g)$ bands $\epsilon_{d\vek\downarrow}$ in the spin minority channel.
At energy larger then $E_0$ the spectral intensity is strongly reduced and eventually it goes up again at even larger energy. The position of the $E_0$ peak shifts from 3.3 eV (\gw@PBE+BSE), to 3.6 eV (PBE0+TDPBE0), and to 4.1 eV (\gw@PBE0+BSE). The $E_0$ energy is always lower compared to the $(\epsilon_{d\vek\downarrow}-\epsilon_{p\vek\downarrow})$ energy difference. We define its binding energy $b_{E_0}=\min_\vek(\epsilon_{d\vek\downarrow}-\epsilon_{p\vek\downarrow})-E_0$. $b_{E_0}\approx1$ eV in both \gw@PBE+BSE (1.1 eV) and PBE0+TDPBE0 (0.9 eV), while $b_{E_0}= 2.4$ eV for \gw@PBE0+BSE.
This discrepancy is again due to the underestimated RPA screening obtained using the PBE0 starting point. This leads to an overestimation of the band-gap which is here compensated by an overestimation of the binding energy and results in a peak position not very different from \gw@PBE+BSE and PBE0+TDPBE0. As for the band structure case, the conclusion here is that it is better to directly compare \gw@PBE+BSE to PBE0+TDPBE0. 

The qualitative behaviour of the spectrum is instead significantly different at energy below $E_0$. All methods show poles at energy well below $E_0$, due to transitions between occupied and flat bands in the minority spin channel, which we identify as $d(t_{2g})$ states (see Supplemental Material \cite{supmat}), and the empty spin minority $d(e_g)$-states. However, within \gw@PBE+BSE, this leads to a series of three bright excitons to appear, which we label $D_i$ in Fig.~\ref{fig:BSE} following the choice done in Ref.~\cite{Jiang2019}. These peaks are located at 1.67 eV ($D_1$), 1.98 eV ($D_2$), 2.21 eV ($D_3$). As already pointed out, the description of the occupied $t_{2g}$ states is questionable in PBE, and so are the absorption features related to them. Indeed the series of the $D$ peaks is missing within \gw@PBE0+BSE and PBE0+TD-BPE0, where all $d(t_{2g}) \rightarrow d(e_g)$ transitions are dark. This is probably due to the fact that the PBE starting point mixes $d$ and $p$ bands character, while PBE0 does not.

Beside these aspects, we are here interested in discussing whether or not NiBr$_2$ is an HEI. We thus need to turn our attention to the lowest energy solutions among the $d(t_{2g}) \rightarrow d(e_g)$ transitions. As discussed, the occupied $d(t_{2g})$ states are very close to the valence band maximum within \gw@PBE, while they are deeper in energy within PBE0 and \gw@PBE0.
As a consequence the \gw@PBE+BSE approach produces negative eigenvalues. There is a doubly degenerate dark exciton at -0.17 eV ($X_{1,1}$) and a non-degenerate exciton at -0.16 eV ($X_{1,2}$). These two excitons share a similar physics and are labeled as $X_1$ in Fig.~\ref{fig:BSE}. Other two dark excitons located at 0.59 ($X_{2,1}$) and 0.68 eV ($X_{2,2}$) are indicated with $X_2$. Both $X_1$ and $X_2$ are associated with transitions between $t_{2g}$ and $e_g$ bands in the minority channel (see Supplemental Material \cite{supmat}).
The $X_1$ poles in particular are the ones which were used to claim the possible existence of an HEI ground state in Ref.~\cite{Jiang2019}.
Instead, within \gw@PBE0+BSE and PBE0+TD-BPE0 no negative eigenvalues appear and therefore there is no EI phase.
Since the qualitative behaviour of these three approaches is the same, and, as discussed in the introduction, the existence of a possible EI phase is mostly quantitative, we have now to address the question: which result is the most reliable?
Both \gw@DFT+BSE schemes are based on a one-shot \gw calculation and plagued by inconsistencies between the approximation used for the ground state and the approximation used for the excited states.
This points to the fact that, as far as this quantitative question is concerned, PBE0+TDPBE0 is the scheme to rely on. For NiBr$_2$ it should be noticed that the \gw@DFT+BSE inconsistency is not so drastic. Indeed
the physics of both the ground state and the excited states is  determined by the same screened interaction $W$.
However, the state-of-the-art \gw@PBE+BSE is affected by a bad description of the occupied $d$ bands in the current study, which is here crucial to address this quantitative answer. \gw@PBE0+BSE suffers from a PBE0 screening which is likely underestimated, although errors induced in the band structure are cancelled by the overestimation of the binding energy, thus affecting in a less severe way the final position of the BSE peaks.
We also notice that in Ref.~\cite{Jiang2019}, the HEI phase is also supported by HSE0+BSE simulations where, however, the inconsistency between ground state and excited states approximation is more problematic. 

In conclusion, our analysis of the band structure and the of the optical spectral features indicates that PBE0+TDPBE0 gives a good overall description of NiBr$_2$. Since this scheme does not produce any negative-energy pole in the optical absorption, we conclude that NiBr$_2$ is not an HEI. This is the most important result of this work.

\section{Conclusions and perspectives \label{Sec:Conclusions}}
In this work we addressed the study of excitonic insulator phases within common \textit{ab initio} theoretical approaches such as time-dependent density functional theory and the Bethe-Salpeter equation of Many-Body Perturbation Theory. We discussed the fact that in order to correctly capture this phase a theoretical approach  should be able to describe two important features, i.e. an electronic-driven broken symmetry and the long range electron-hole interaction. That is why TDDFT within the common LDA/GGA approximations (which miss the long range electron-hole interaction) cannot correctly describe this phase. The BSE is more appropriate, but, in order to correctly describe an electronic-driven broken symmetry, it should be used within consistent approximations. We discussed that this corresponds to use the same level of approximation for the self-energy used in a self-consistent ground-state calculation and for the kernel of the BSE.

For the specific case of ferromagnetic NiBr$_2$,
by comparing the optical spectrum and the energy position of the lowest energy dark excitons obtained with different schemes, we arrived at the conclusion that the PBE0+TDPBE0 scheme is the most reliable, also because it uses the same level of approximation for the ground state and for the excited states.
Within PBE0+TDPBE0 all dark poles have a positive energy, which indicates that NiBr$_2$ is not an HEI, contrary to previous assignments. 
Moreover, we find that the occurrence of negative energy peaks in the \gw@PBE+BSE is due to the starting point dependence of the \gw\, scheme, rather than to the inconsistency between ground-state and excited state approximations. 
Indeed, the PBE starting point gives a questionable description of the $t_{2g}$ states, which leads,  within \gw@PBE, to $t_{2g}$ bands located too high with respect to the occupied $p$ states. This does not happen within PBE0 and \gw@PBE0 (and also with respect to other approximations studied in the literature such as HSE0 and PBE+$U$). 

More in general, our study points to the fact that using consistent approximations in \textit{ab initio} theories is of crucial importance in order to have reliable results. 

\begin{acknowledgments}
S. Di Sabatino and P. Romaniello would like to acknowledge financial support by the EUR grant NanoX ANR-17-EURE-0009 in the framework of the ``Programme des Investissements d'Avenir" and by the ANR (project ANR-18-CE30-0025 and ANR-19-CE30-0011). A. M.-S. acknowledge financial support by Ram\'on y Cajal programme (grant RYC2018-024024-I; MINECO, Spain), Agencia Estatal de Investigación (AEI), through the project PID2020-112507GB-I00 (Novel quantum states in heterostructures of 2D materials), and Generalitat Valenciana, program SEJIGENT (reference 2021/034), project Magnons in magnetic 2D materials for a novel electronics (2D MAGNONICS), and Planes complementarios de I+D+I en materiales avanzados, project SPINO2D, reference MFA/2022/009.

\end{acknowledgments}

\end{document}


\title{Supplemental Material to ``Assignment of excitonic insulators in \textit{ab initio} theories: the case of NiBr$_2$"}
\author{Stefano Di Sabatino}
\email{stefano.disabatino@irsamc.ups-tlse.fr}
\affiliation{\lpt}
\author{Alejandro Molina-S\'{a}nchez}
\email{alejandro.molina@uv.es}
\affiliation{\al}
\author{Pina Romaniello}
\email{pina.romaniello@irsamc.ups-tlse.fr}
\affiliation{\lpt}
\author{Davide Sangalli}
\email{davide.sangalli@ism.cnr.it}
\affiliation{\san}

\maketitle 

\begin{figure}[!b]
    \centering
    \includegraphics[width=0.45\textwidth]{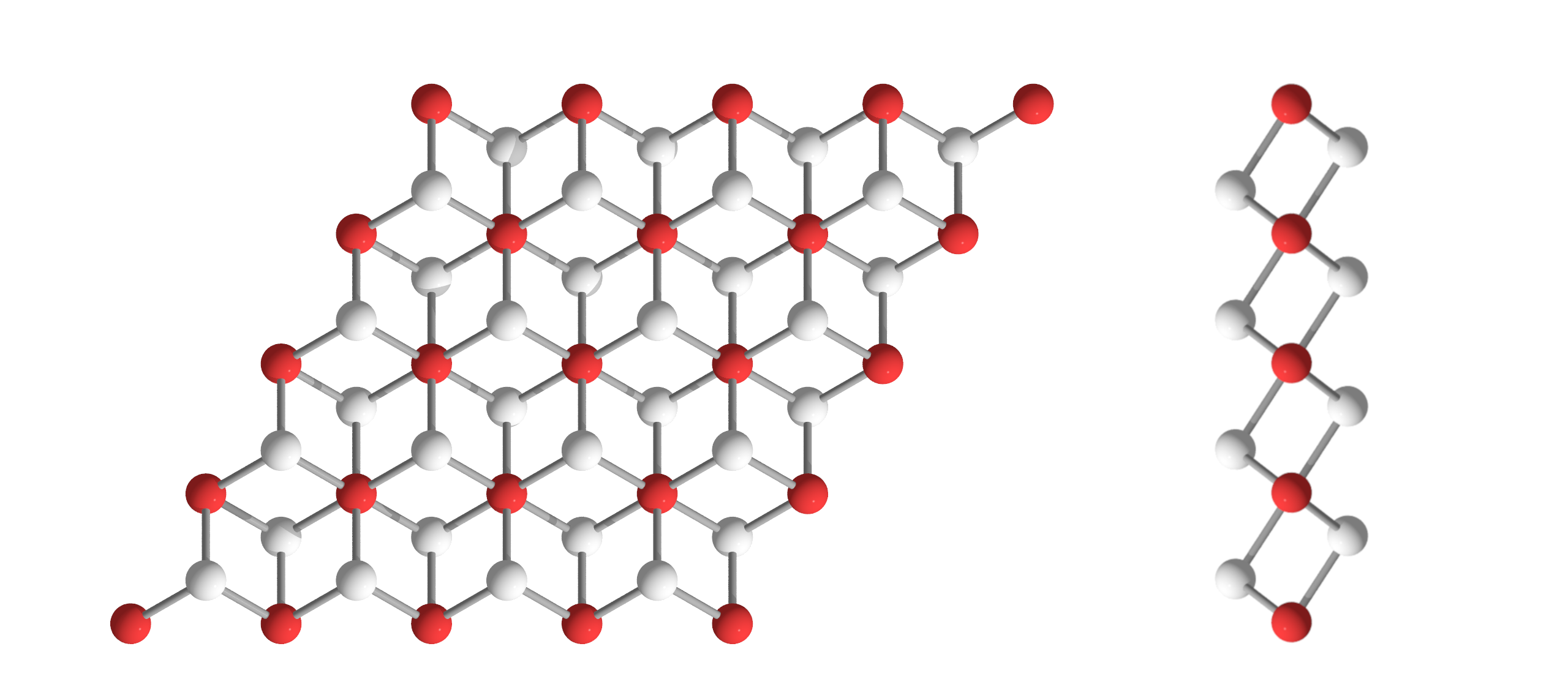}
    \caption{NiBr$_2$ monolayer structure. Ni atoms are indicated in red, Br atoms in white.}
    \label{fig:struct}
\end{figure}

\section{\gw@PBE convergence\label{gwconv}}

\begin{figure}
    \centering
    \includegraphics[width=0.33\textwidth]{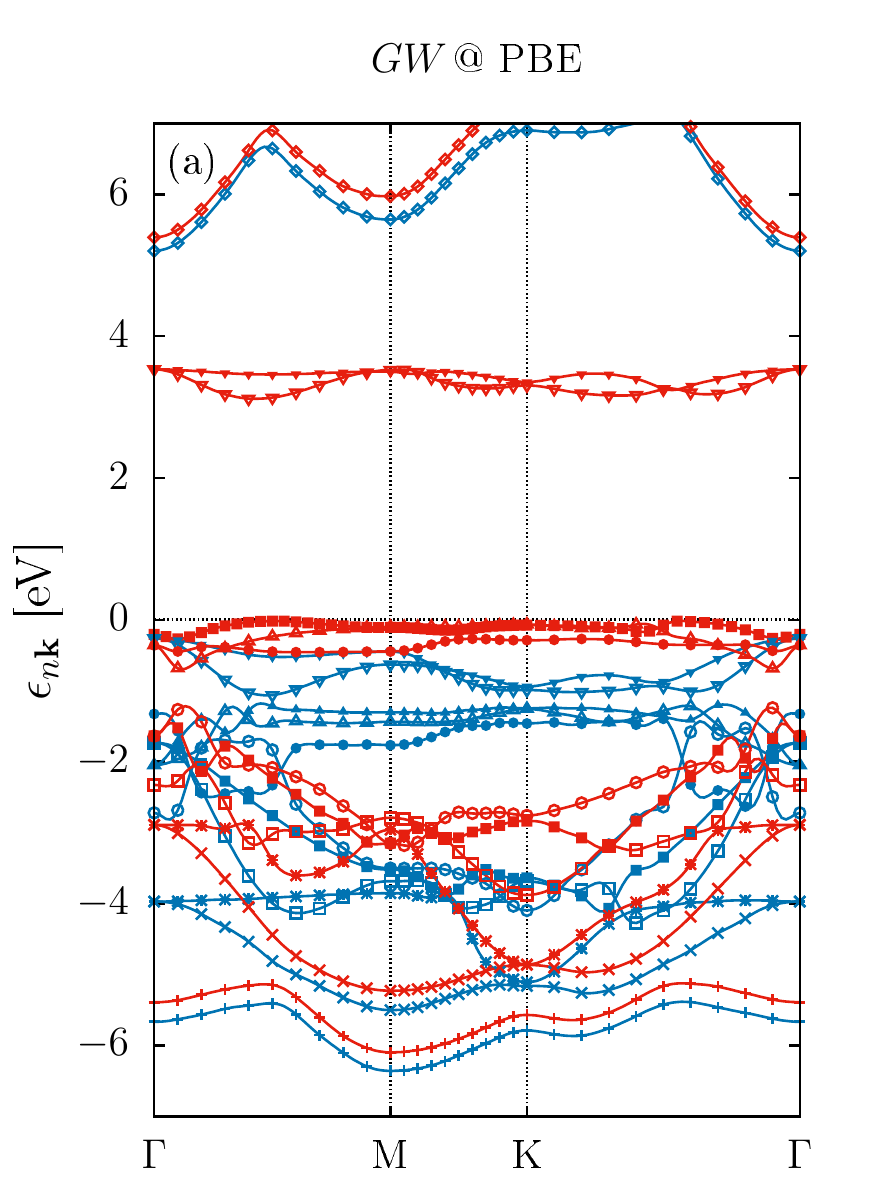}
    \includegraphics[width=0.33\textwidth]{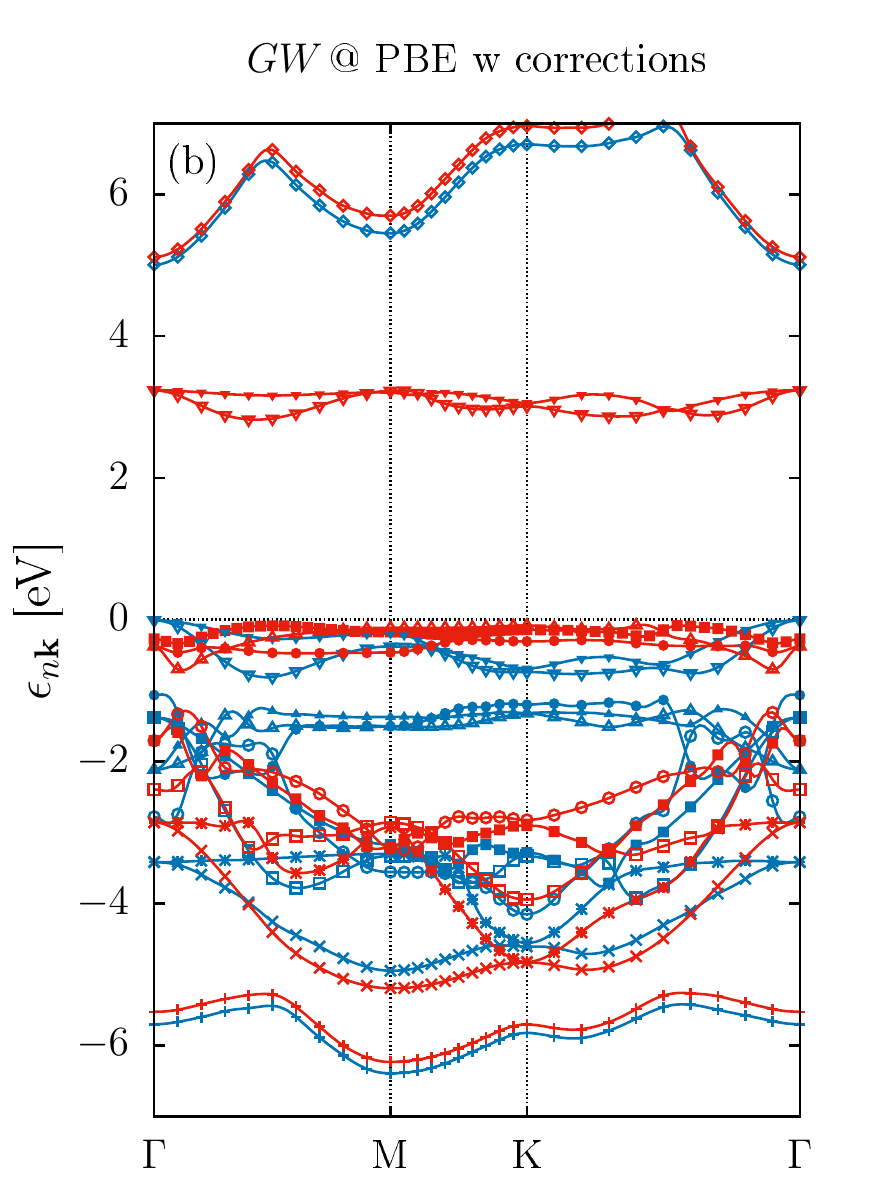} 
    \caption{Convergence of the QP \gw corrections. (a) Band structure using a 19$\times$19$\times$1 $\vect{k}$-point grid, an energy cutoff of $E_{\text{cut}}$=10 Ry for the screening and the correlation parts of the self-energy, $N_b=200$ bands in the screening, and $500$ bands for the evaluation of the self-energy. Symbols indicate the different bands considered. (b) Band structure after the introduction of the corrections obtained from the extrapolation procedure. \label{fig:convGWextra1}}
\end{figure}

\begin{figure}
    \includegraphics[width=0.28\textwidth]{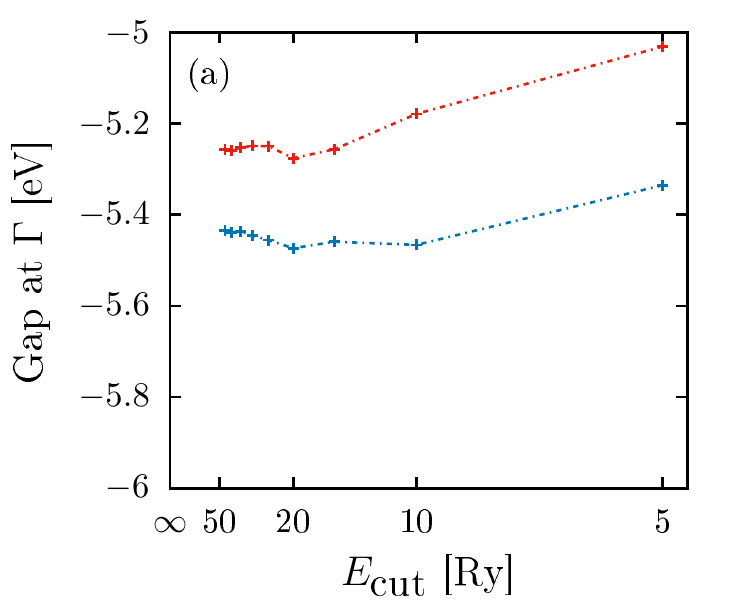}
    \includegraphics[width=0.28\textwidth]{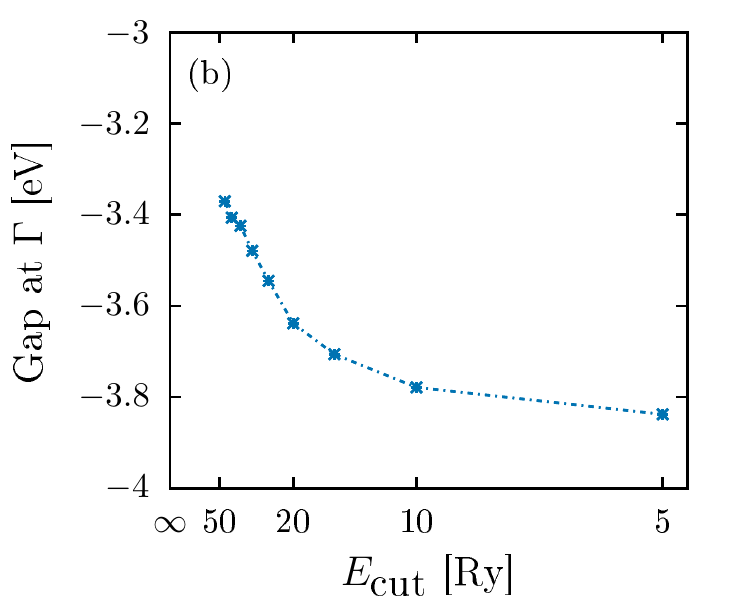}
    \includegraphics[width=0.28\textwidth]{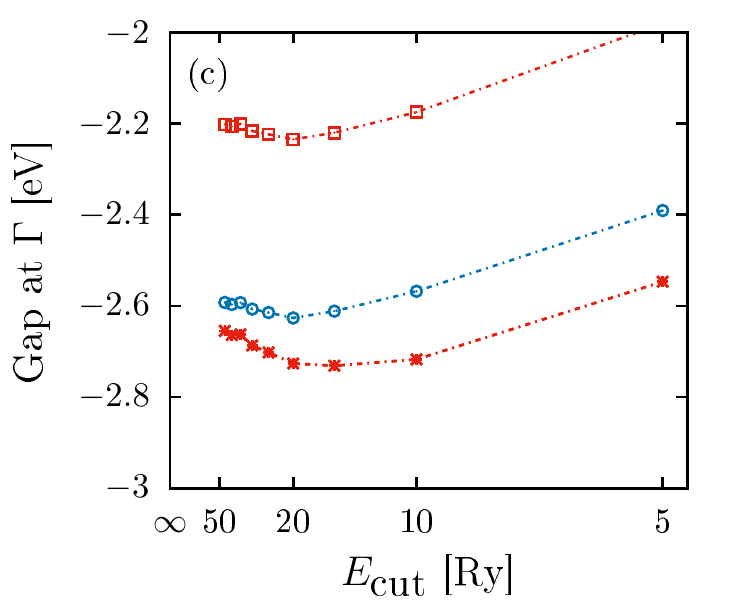}
    \includegraphics[width=0.28\textwidth]{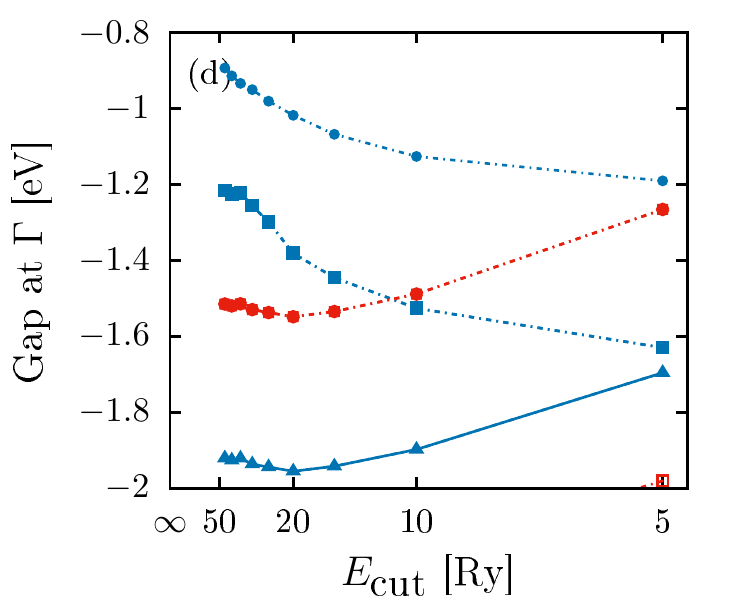}
    \includegraphics[width=0.28\textwidth]{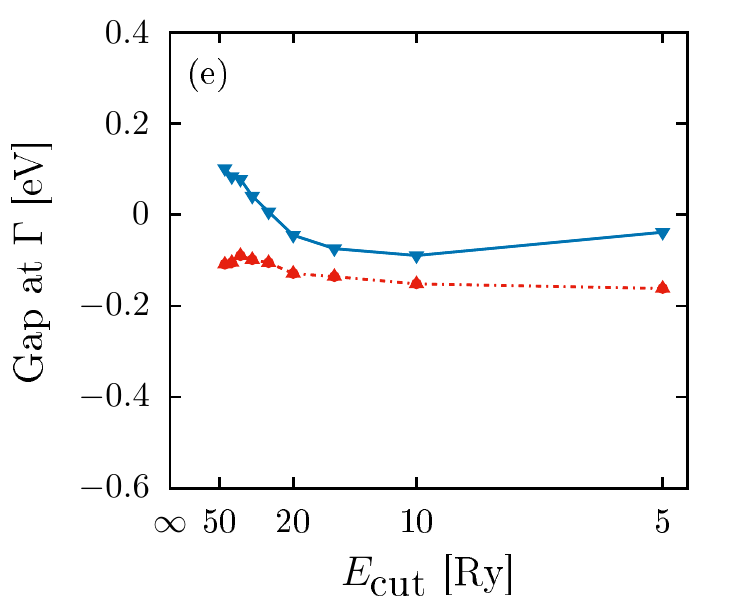}
    \includegraphics[width=0.28\textwidth]{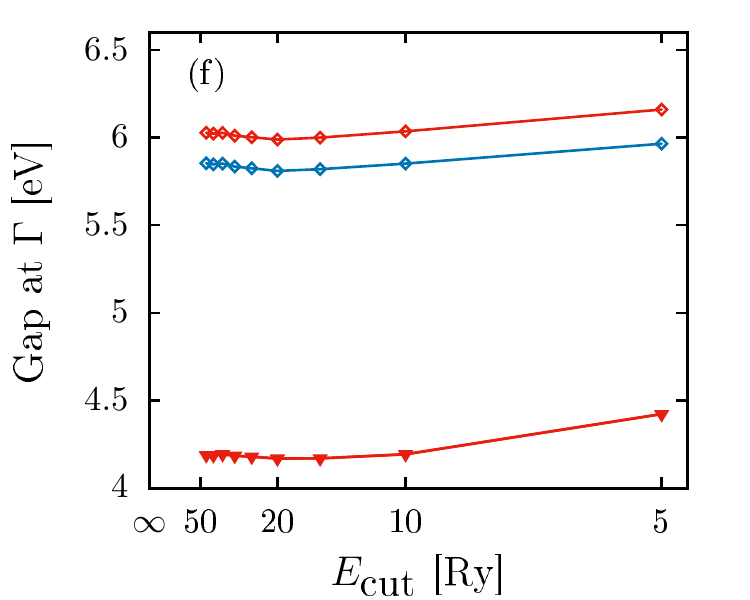}
    \includegraphics[width=0.28\textwidth]{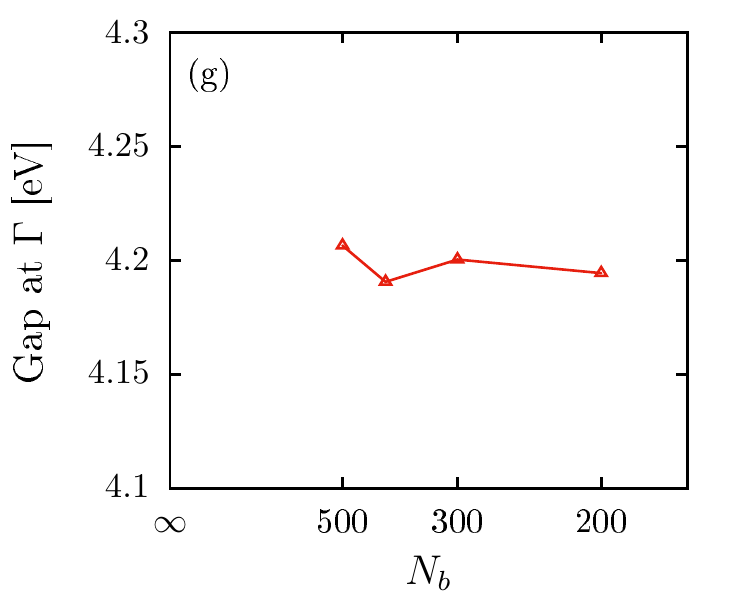}
    \includegraphics[width=0.28\textwidth]{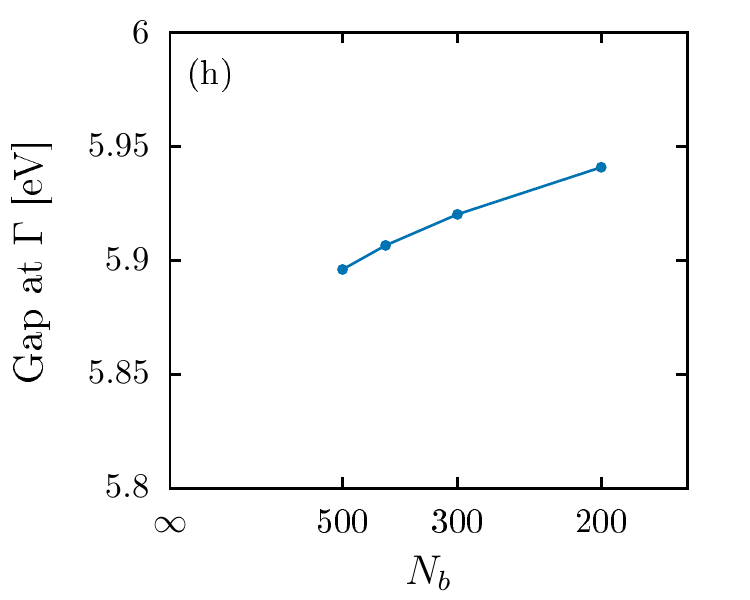}
    \includegraphics[width=0.28\textwidth]{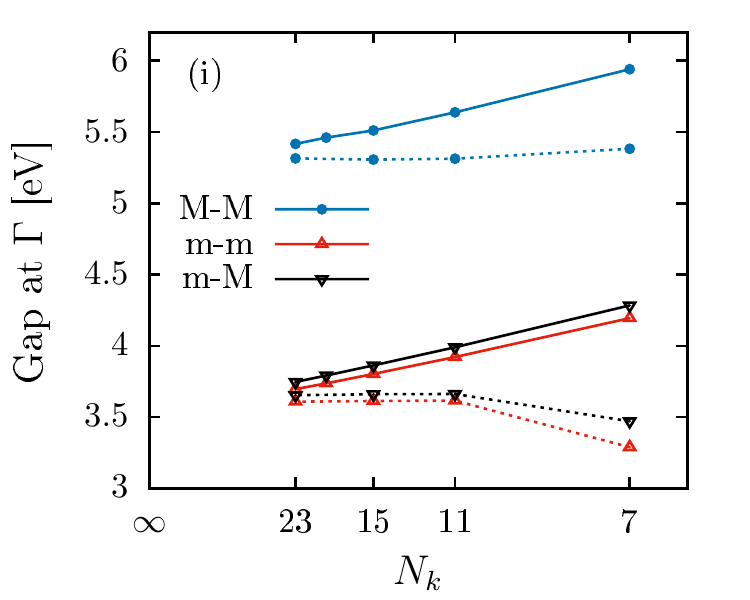}
    \caption{
    (c)-(h) Convergence of the difference $\epsilon_{n}-\epsilon_{\text{VBM}}$ at the $\Gamma$ point with respect to the energy cutoff $E_{\text{cut}}$ in the screening and the correlation part of the self-energy. The convergence is studied with a $7\times 7\times 1$ $\vect{k}$-point grid. Some bands in the majority spin channel show a slower convergence, with a $1/E_{\text{cut}}$ behaviour. For these bands an extrapolation procedure can be used.
    The effect is an inversion of the top valence band at $\Gamma$.
    (i)-(j) Convergence of the m-m, M-M gaps at $\Gamma$ with respect to the number of bands $N_b$ in the screening. 
    (k) Convergence of the M-M, m-m and m-M gaps at $\Gamma$ with respect to the number of $\vect{k}$ points. Results obtained with the random integration scheme proposed in Ref.~\onlinecite{guandalini2022efficient} are also presented with a dashed line for comparison. 
}
    \label{fig:convGWextra2}
\end{figure}

For monolayer NiBr$_2$, whose structure is shown in Fig. \ref{fig:struct},  particular attention has to be paid when converging the \gw quasiparticle corrections. 
In Fig.~\ref{fig:convGWextra1}(a) we report the band structure within \gw@PBE using the following parameters: a 19$\times$19$\times$1 $\vect{k}$-point grid, an energy cutoff of $E_{\text{cut}}$=10 Ry  for the screening and the correlation parts of the self-energy $\Sigma_c$, and $N_b=200$ bands in the screening.
In Fig.~\ref{fig:convGWextra1}(b) we report the same band structure after the introduction of the correction obtained from the extrapolation process described in the following.
In Fig.~\ref{fig:convGWextra2}(a)-(f) we report the difference between $\epsilon_{n\vect{k}}$ at $\Gamma$ (See Fig.~\ref{fig:convGWextra1}(a)) and the top valence band at 
$\Gamma$ as function of the energy cutoff in the screening.
We notice that some bands in the spin up channel converge quite slowly with respect to the energy cutoff, showing a $\sim 1/E_{\text{cut}}$ behaviour. 
The $E_{\text{cut}}\rightarrow\infty$ limit can be extrapolated from the finite cutoff data.
As can be seen from the band structure in Fig.~\ref{fig:convGWextra1}(a) the VBM is in the minority channel at a point about half way between the $\Gamma$ and M symmetry points. However increasing the energy cutoff we observe a shift between the $t_{2g}$ in the minority channel and the $e_g$ in the majority channel. This produces an inversion of the channel of the VBM, which is now located at the $\Gamma$ point. Results qualitatively similar to our unconverged band structure are also presented in Ref.~\cite{c2db}, and in Ref.~\cite{Jiang2019} for the analogous case of NiCl$_2$.
The minority-minority spin gap, instead, does not pose any particular problem; at $E_{\text{cut}}=10$ Ry the gap is converged by less than 0.1 eV.
In Fig.~\ref{fig:convGWextra2}(g)-(h) we report the convergence of the M-M and m-m direct gap at $\Gamma$ with respect to the number of bands in the screening. The convergence error for both gaps is less than 0.1 eV when using $N_b=200$.
In Fig.~\ref{fig:convGWextra2}(i) we report the convergence of the M-M, m-m and m-M direct band gaps at $\Gamma$ with respect to the $\vect{k}$-point grid. No significant difference is observed between the two different spin channels.
When using a $19\times19\times1$ $\vect{k}$-point grid, the correction due to the convergence with respect to the grid corresponds to a rigid shift of the conduction bands with respect to the valence bands of about -0.24 eV.
For comparison we report the band gap obtained using 
the stochastic integration scheme for the screened potential described in Ref.~\onlinecite{guandalini2022efficient} in order to speed up the convergence with respect to the number of $\vect{k}$ points.

\clearpage 

\section{Excitons analysis\label{excanalysis}}

We present here a detailed analysis of the excitonic wave-functions of the dark energy states for GW@PBE+BSE, GW@PBE0+BSE, and PBE0+TD-PBE0.

The atomic lattice used for NiBr$_2$ respects the $D_{3d}$ space group with 12 symmetry operation divided into 6 irreducible representations. Four representations of dimension 1 ($A_{1g}$, $A_{2g}$, $A_{1u}$, and $A_{2u}$) and two representations of dimension 2 ($E_g$ and $E_u$). Bright excitons belong either to the $E_u$ (in plane bright) or the $A_{2u}$ (out of plane bright). We remark that only negative poles belonging to the $E_g$ (dark excitons) or $E_u$ (bright excitons) groups can lead to an excitonic insulator phase according to the discussion in the main text.

\begin{figure*}[!b]
    \centering
    \includegraphics[width=0.80\textwidth]{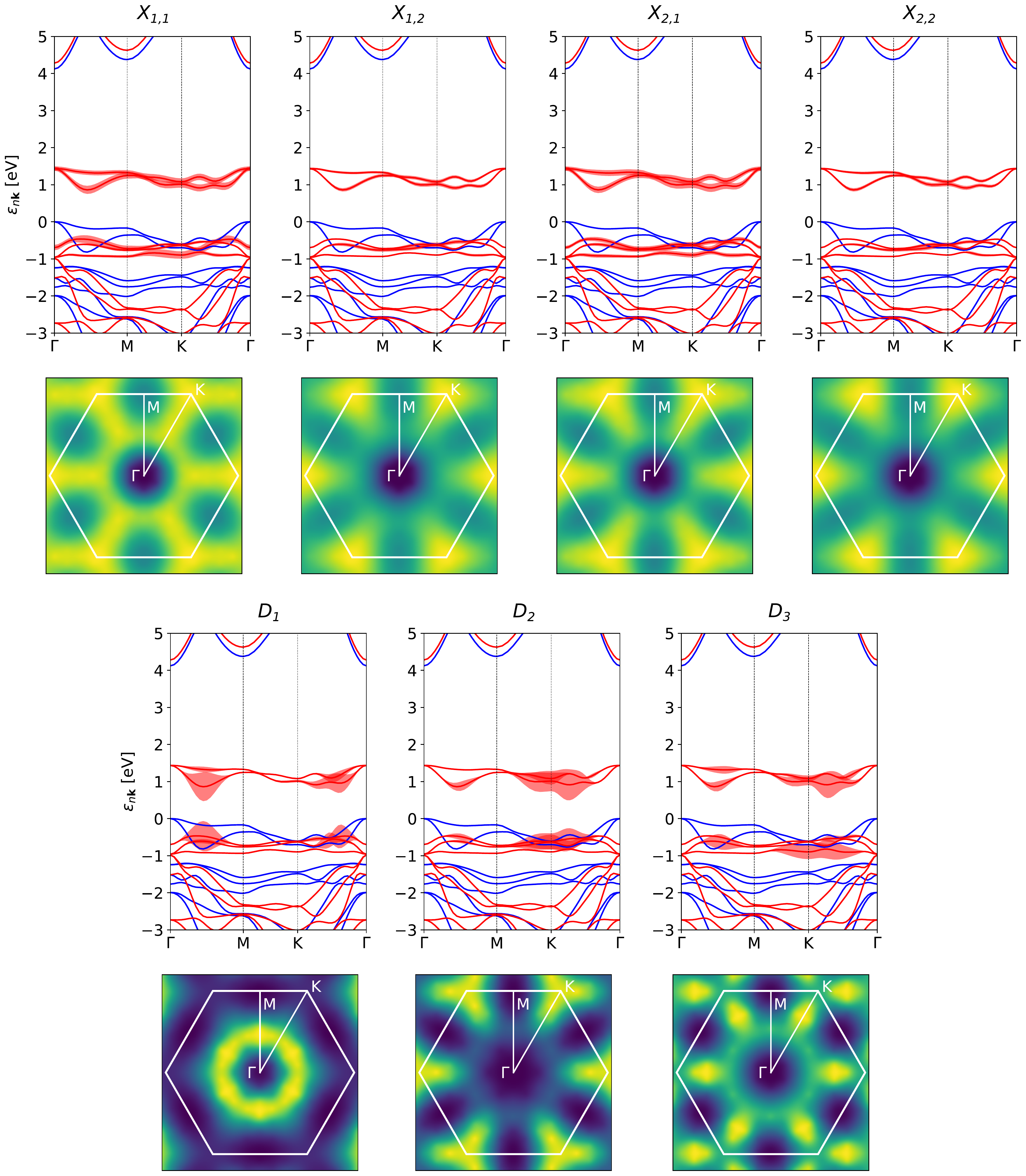}
    \caption{\gw@PBE+BSE excitonic wavefunction represented on the PBE band structure (upper panels), and in reciprocal space (lower panels). We used the PBE bands for simplicity, since $GW$ only introduces an energy shift without changing the band character.  First 4 dark excitons ($X_{1,1}$, $X_{1,2}$, $X_{2,1}$, $X_{2,2}$) and 3 bright excitons ($D_1$, $D_2$, $D_3$).}
    \label{fig:BSEExc}
\end{figure*}

Within \gw@PBE+BSE the dark excitons are located at
 -0.17 eV ($X_{1,1}$, doubly degenerate)
 -0.16 eV ($X_{1,2}$, non degenerate) 
  0.59 eV ($X_{2,1}$, doubly degenerate)
  0.68 eV ($X_{2,2}$, non degenerate)
plus three all doubly degenerate bright excitons
 at $1.67$ eV ($D_1$),
 $1.98$ eV ($D_2$), and
 $2.21$ eV ($D_3$).
 In Fig. \ref{fig:BSEExc} we present the wave function of the excitons here considered with the band transitions involved. Both the dark excitons $X_1$, $X_2$ and the bright excitons $D_i$  are related to $t_{2g} \rightarrow e_g$.

\begin{figure*}
    \centering
    \includegraphics[width=0.99\textwidth]{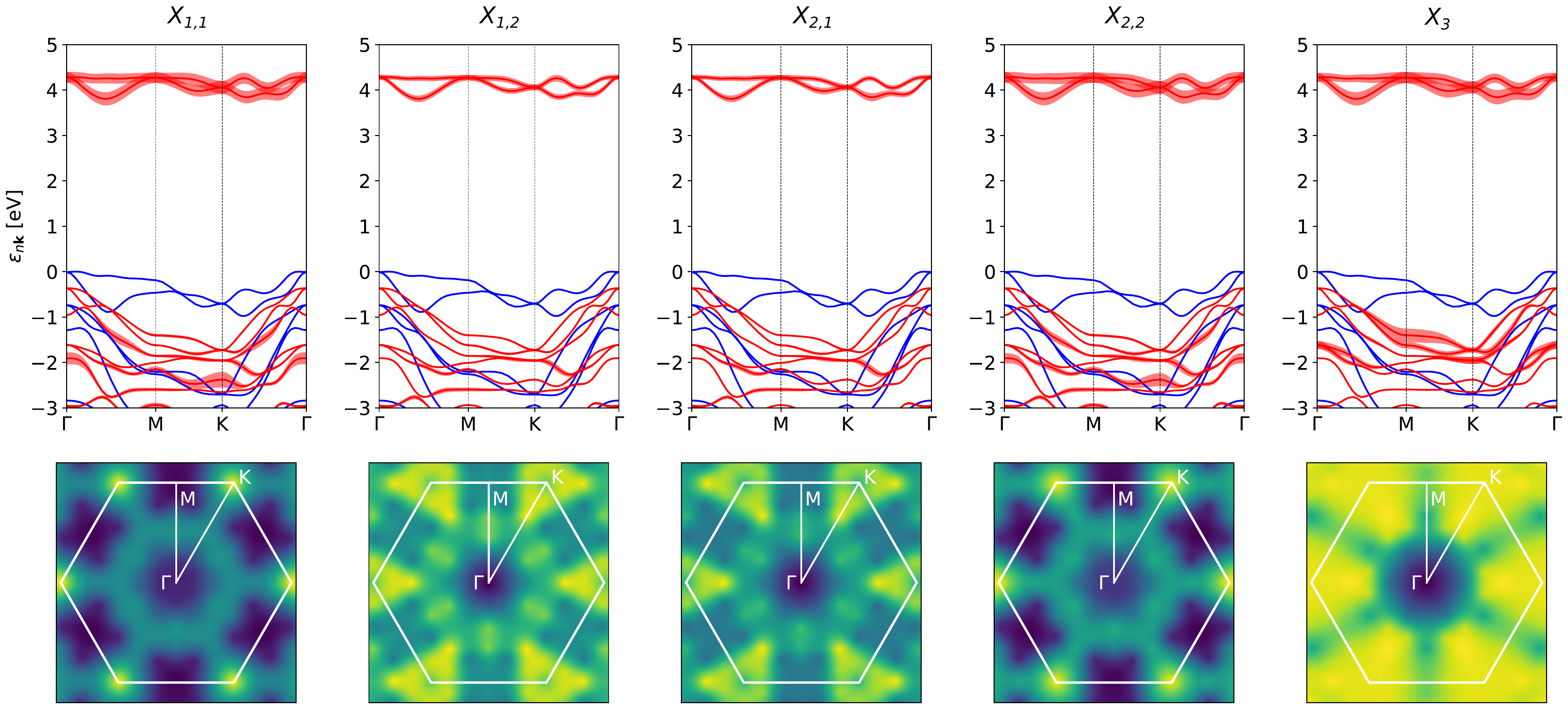}
    \caption{\gw@PBE0+BSE excitonic wavefunction represented on the PBE0 band structure (upper panels), and in reciprocal space (lower panels). We used the PBE0 bands for simplicity, since $GW$ only introduces an energy shift without changing the band character.  First 5 dark excitons ($X_{1,1}$, $X_{1,2}$, $X_{2,1}$, $X_{2,2}$,  $X_{3}$).}
    \label{fig:BSEExcPBE0}
\end{figure*}

Within \gw@PBE0+BSE the dark excitons are located at
 1.45 eV ($X_{1,1}$, doubly degenerate),       
 1.51 eV ($X_{1,2}$, non-degenerate),        
 2.00 eV ($X_{2,1}$, non-degenerate),      
 2.05 eV ($X_{2,2}$, doubly degenerate), and    3.99 eV ($X_{3}$, doubly degenerate).
Excitonic wave functions and band transition involved are reported in Fig. \ref{fig:BSEExcPBE0}

\begin{figure*}
    \centering
    \includegraphics[width=0.82\textwidth]{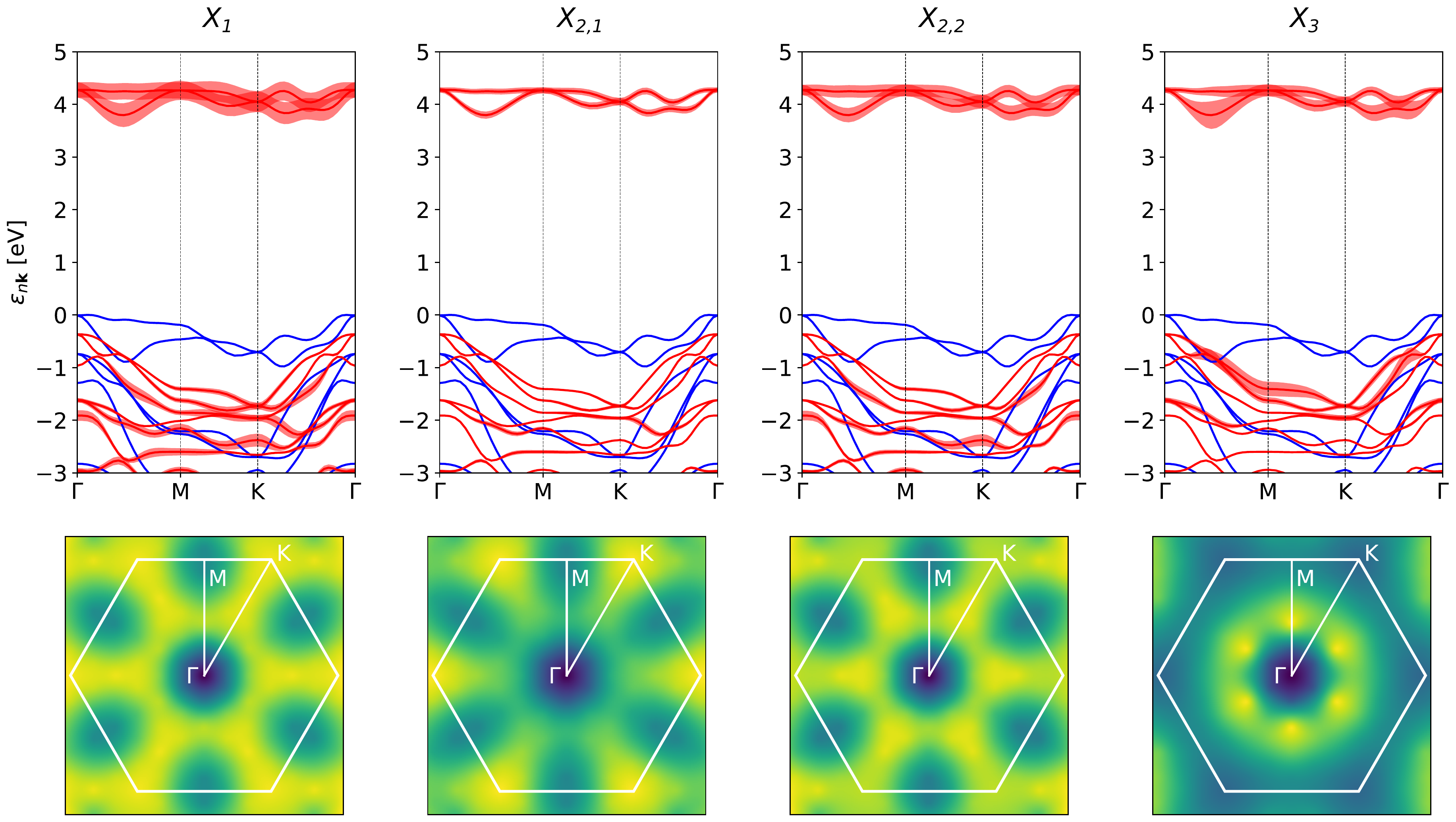}
    \caption{PBE0+TDPBE0 excitonic wavefunction represented on the PBE0 band structure (upper panels), and in reciprocal space (lower panels). First 4 dark excitons ($X_{1}$, $X_{2,1}$, $X_{2,2}$,  $X_{3}$).}
    \label{fig:BSEExcTD}
\end{figure*}

Within TD-PBE0 we obtain dark excitons at
 1.07 eV ($X_{1}$, triply degenerate), 
 1.31 eV ($X_{2,1}$, doubly degenerate),
 1.34 eV ($X_{2,2}$, nondegenerate), and
 3.55 eV ($X_{3}$, doubly degenerate). 
Excitonic wave functions and band transitions involved are reported in Fig. \ref{fig:BSEExcTD}.